\documentclass[journal=jctcce]{achemso}
\setkeys{acs}{articletitle = true}

\usepackage{graphicx,dcolumn,bm,xcolor,hyperref,multirow,amsmath,amssymb,amsfonts,physics}
\usepackage[normalem]{ulem}
\usepackage[version=4]{mhchem}
\usepackage[compat=1.1.0]{tikz-feynman}
\usepackage[tracking]{microtype}
\usepackage{mathpazo,libertine}

\definecolor{mypink}{rgb}{0.858, 0.188, 0.478}

\newcommand{\titou}[1]{\textcolor{black}{#1}}

\newcommand{\evGW}{evGW}	
\newcommand{\qsGW}{qsGW}	
\newcommand{\GOWO}{G$_0$W$_0$}	
\newcommand{\GW}{GW}		
\newcommand{\GnWn}[1]{G$_{#1}$W$_{#1}$}		

\newcommand{\Egap}{E_\text{gap}}

\newcommand{\RH}{R_{\ce{H2}}}
\newcommand{\RF}{R_{\ce{F2}}}
\newcommand{\RBeO}{R_{\ce{BeO}}}

\newcommand{\nDIIS}{N^\text{DIIS}}
\newcommand{\maxDIIS}{N_\text{max}^\text{DIIS}}

\newcommand{\e}[1]{\epsilon_{#1}}
\newcommand{\eHF}[1]{\epsilon^\text{HF}_{#1}}

\newcommand{\eGOWO}[1]{\epsilon^\text{\GOWO}_{#1}}
\newcommand{\eGW}[1]{\epsilon^\text{\GW}_{#1}}

\newcommand{\Om}[1]{\Omega_{#1}}
\newcommand{\eHOMO}{\epsilon_\text{HOMO}}
\newcommand{\eLUMO}{\epsilon_\text{LUMO}}
\newcommand{\HOMO}{\text{HOMO}}
\newcommand{\LUMO}{\text{LUMO}}

\newcommand{\SigC}[1]{\Sigma^\text{c}_{#1}}
\newcommand{\SigCp}[1]{\Sigma^\text{p}_{#1}}
\newcommand{\SigCh}[1]{\Sigma^\text{h}_{#1}}

\newcommand{\Z}[1]{Z_{#1}}

\newcommand{\beGnWn}[1]{\boldsymbol{\epsilon}^\text{\GnWn{#1}}}
\newcommand{\bdeGnWn}[1]{\Delta\boldsymbol{\epsilon}^\text{\GnWn{#1}}}

\newcommand{\bOm}{\boldsymbol{\Omega}}
\newcommand{\bA}{\boldsymbol{A}}
\newcommand{\bB}{\boldsymbol{B}}
\newcommand{\bX}{\boldsymbol{X}}
\newcommand{\bY}{\boldsymbol{Y}}

\newcommand{\lcpq}{Laboratoire de Chimie et Physique Quantiques, Universit\'e de Toulouse, CNRS, UPS, France}
\newcommand{\lpt}{Laboratoire de Physique Th\'eorique, Universit\'e de Toulouse, CNRS, UPS, France}
\newcommand{\etsf}{European Theoretical Spectroscopy Facility (ETSF)}

\title{Unphysical Discontinuities in GW Methods}

\author{Micka\"el V\'eril}
\affiliation{\lcpq}
\author{Pina Romaniello}
\affiliation{\lpt}
\altaffiliation{\etsf}
\author{J.~A.~Berger}
\affiliation{\lcpq}
\altaffiliation{\etsf}	
\author{Pierre-Fran\c{c}ois Loos}
\affiliation{\lcpq}
\email{loos@irsamc.ups-tlse.fr}

\keywords{many-body perturbation theory; multiple solutions; GW approximation; self-consistent scheme}

\begin{document}	

\begin{abstract}
We report unphysical irregularities and discontinuities in some key experimentally-measurable quantities computed within the GW approximation 
of many-body perturbation theory applied to molecular systems.
In particular, we show that the solution obtained with partially self-consistent GW schemes depends on the algorithm one uses to solve self-consistently the quasi-particle (QP) equation.
The main observation of the present study is that each branch of the self-energy is associated with a distinct QP solution, and that each switch between solutions implies a significant discontinuity in the quasiparticle energy as a function of the internuclear distance.
Moreover, we clearly observe ``ripple'' effects, i.e., a discontinuity in one of the QP energies induces (smaller) discontinuities in the other QP energies.  
Going from one branch to another implies a transfer of weight between two solutions of the QP equation.
The case of occupied, virtual and frontier orbitals are separately discussed on distinct diatomics.
In particular, we show that multisolution behavior in frontier orbitals is more likely if the HOMO-LUMO gap is small.
\end{abstract}

\maketitle

\section{Background}
Many-body perturbation theory methods based on the one-body Green function $G$ are fascinating as they are able to transform an unsolvable many-electron problem into a set of non-linear one-electron equations, thanks to the introduction of an effective potential $\Sigma$, the self-energy.
Electron correlation is explicitly incorporated via a sequence of self-consistent steps connected by Hedin's equations. \cite{Hedin_1965}
In particular, Hedin's approach uses a dynamically screened Coulomb interaction $W$ instead of the standard bare Coulomb interaction.
Important experimental properties such as ionization potentials, electron affinities as well as spectral functions, which are related to 
direct and inverse photo-emission, can be obtained directly from the one-body Green function. \cite{Onida_2002}
A particularly successful and practical approximation to Hedin's equations is the so-called GW approximation \cite{Onida_2002, Aryasetiawan_1998, Reining_2017} which bypasses the calculation of the most complicated part of Hedin's equations, the vertex function. \cite{Hedin_1965}

Although (perturbative) {\GOWO} is probably the simplest and most widely used GW variant, \cite{Hybertsen_1985a, vanSetten_2013, Bruneval_2012, Bruneval_2013, vanSetten_2015, vanSetten_2018} its starting point dependence has motivated the development of partially \cite{Hybertsen_1986, Shishkin_2007, Blase_2011, Faber_2011, Faleev_2004, vanSchilfgaarde_2006, Kotani_2007, Ke_2011, Kaplan_2016} and fully \cite{Stan_2006, Stan_2009, Rostgaard_2010, Caruso_2012, Caruso_2013, Caruso_2013a, Caruso_2013b, Koval_2014, Wilhelm_2018} self-consistent versions in order to reduce or remove this undesirable feature.
Here, we will focus our attention on partially self-consistent schemes as they have demonstrated comparable accuracy and are computationally lighter than the fully self-consistent version. \cite{Caruso_2016}
Moreover, they are routinely employed for solid-state and molecular calculations and are available in various computational packages. \cite{Blase_2011, Blase_2018, Bruneval_2016, vanSetten_2013, Kaplan_2015, Kaplan_2016, Krause_2017, Caruso_2016, Maggio_2017}
Recently, an ever-increasing number of successful applications of partially self-consistent GW methods have sprung in the physics and chemistry literature for molecular systems, \cite{Ke_2011, Bruneval_2012, Bruneval_2013, Bruneval_2015, Bruneval_2016, Bruneval_2016a, Koval_2014, Hung_2016, Blase_2018, Boulanger_2014, Jacquemin_2017, Li_2017, Hung_2016, Hung_2017, vanSetten_2015, vanSetten_2018} as well as extensive and elaborate benchmark sets. \cite{vanSetten_2015, Maggio_2017, vanSetten_2018, Richard_2016, Gallandi_2016, Knight_2016, Dolgounitcheva_2016, Bruneval_2015, Jacquemin_2015}

There exist two main types of partially self-consistent GW methods: 
i) \textit{``eigenvalue-only quasiparticle''} GW ({\evGW}), \cite{Hybertsen_1986, Shishkin_2007, Blase_2011, Faber_2011} 
where the quasiparticle (QP) energies are updated at each iteration, and 
ii) \textit{``quasiparticle self-consistent''} GW ({\qsGW}), \cite{Faleev_2004, vanSchilfgaarde_2006, Kotani_2007, Ke_2011, Kaplan_2016} 
where one updates both the QP energies and the corresponding orbitals.
Note that a starting point dependence remains in {\evGW} as the orbitals are not self-consistently optimized in this case.

In a recent article, \cite{Loos_2018} while studying a model two-electron system, \cite{Seidl_2007, Loos_2009a, Loos_2009c, Loos_2010e, Loos_2011b, Gill_2012} we have observed that, within partially self-consistent GW (such as {\evGW} and {\qsGW}), one can observe, in the weakly correlated regime, (unphysical) discontinuities in the energy surfaces of several key quantities (ionization potential, electron affinity, HOMO-LUMO gap, total and correlation energies, as well as vertical excitation energies).
In the present manuscript, we provide further evidences and explanations of this undesirable feature in real molecular systems.
For sake of simplicity, the present study is based on simple closed-shell diatomics (\ce{H2}, \ce{F2} and \ce{BeO}).
However, the same phenomenon can be observed in many other molecular systems, such as \ce{LiF}, \ce{HeH+}, \ce{LiH}, \ce{BN}, \ce{O3}, etc.
Although we mainly focus on {\GOWO} and {\evGW}, similar observations can be made in the case of {\qsGW} and second-order Green function (GF2) methods. \cite{SzaboBook, Casida_1989, Casida_1991, Stefanucci_2013, Ortiz_2013, Phillips_2014, Phillips_2015, Rusakov_2014,Rusakov_2016, Hirata_2015, Hirata_2017, Loos_2018}
Unless otherwise stated, all calculations have been performed with our locally-developed GW software, which closely follows the MOLGW implementation. \cite{Bruneval_2016}

\begin{figure*}
	\includegraphics[width=\linewidth]{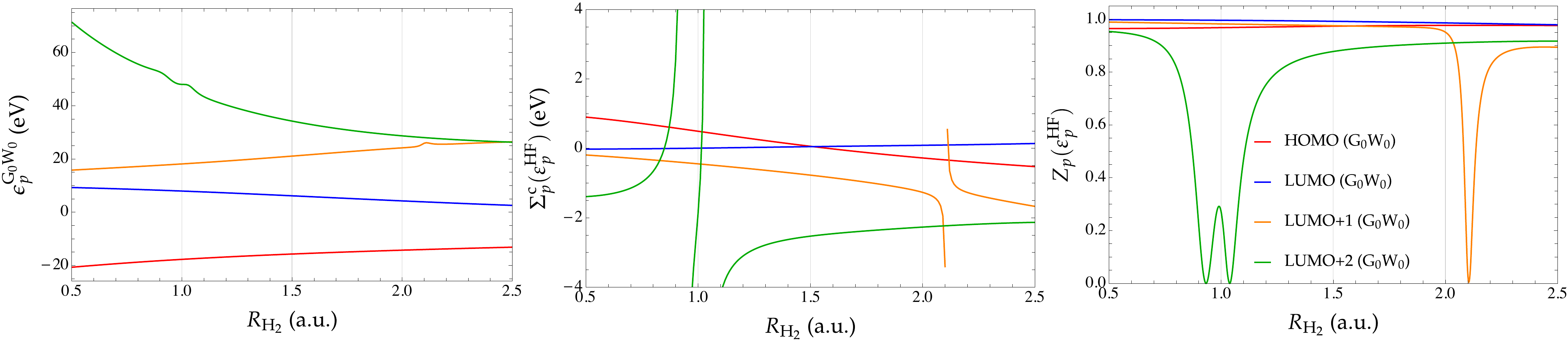}
	\includegraphics[width=\linewidth]{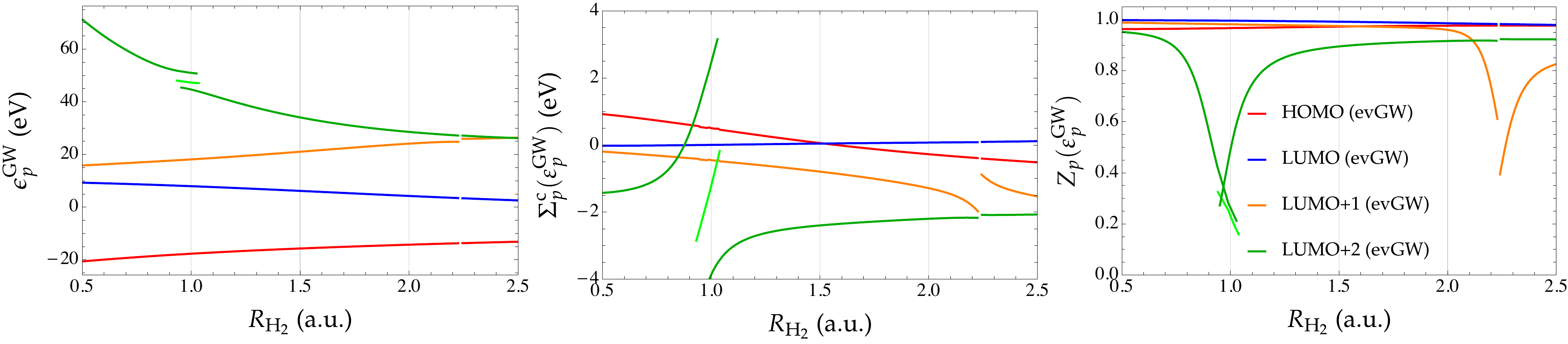}
	\caption{
	\label{fig:H2}
	QP energies (left), correlation part of the self-energy (center) and renormalization factor (right) as functions of the internuclear distance $\RH$ for various orbitals of \ce{H2} at the {\GOWO}@HF/6-31G (top) and {\evGW}@HF/6-31G (bottom) levels.
	For convenience, the intermediate (center) branch is presented in lighter green for the LUMO+2.
}
\end{figure*}	

\begin{figure}
	\includegraphics[width=0.6\linewidth]{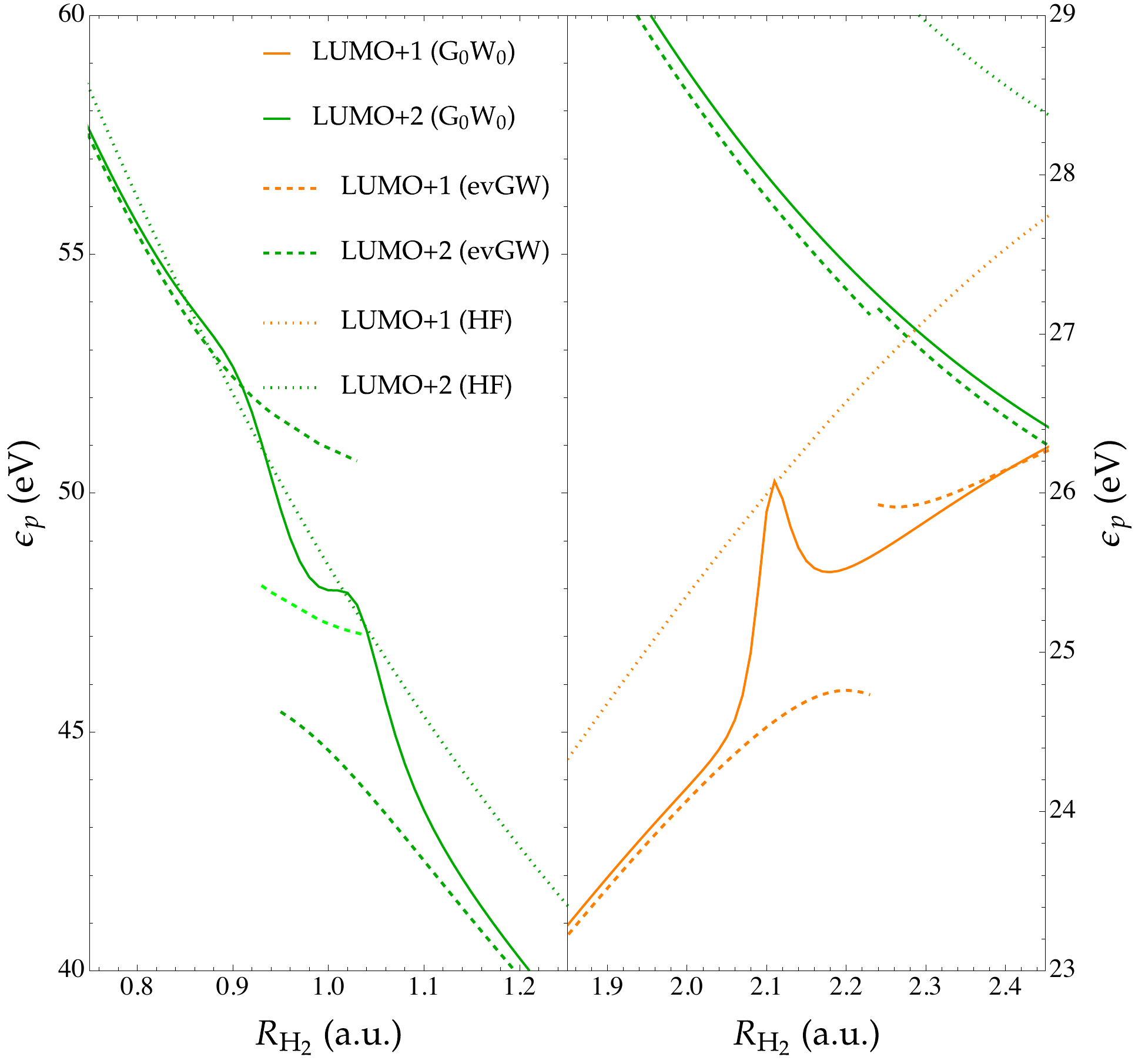}
	\caption{
	\label{fig:H2-zoom}
	HF orbital energies (dotted lines) and QP energies as functions of the internuclear distance $\RH$ for the LUMO+1 and LUMO+2 orbitals of \ce{H2} at the {\GOWO}@HF/6-31G (solid lines) and {\evGW}@HF/6-31G (dashed lines) levels.
	For convenience, the intermediate (center) branch is presented in lighter green for the LUMO+2.
}
\end{figure}	

\section{Theory}
Here, we provide brief details about the main equations and quantities behind {\GOWO} and {\evGW} considering a (restricted) Hartree-Fock (HF) starting point. \cite{SzaboBook} 
More details can be found, for example, in Refs.~\citenum{vanSetten_2013, Kaplan_2016, Bruneval_2016}.

For a given (occupied or virtual) orbital $p$, the correlation part of the self-energy \titou{is conveniently split in its hole (h) and particle (p) contributions}
\begin{equation}
\label{eq:SigC}
	\SigC{p}(\omega) = \SigCp{p}(\omega) + \SigCh{p}(\omega),
\end{equation}
\titou{which, within the GW approximation, read}
\begin{subequations}
\begin{align}
\label{eq:SigCh}
	\SigCh{p}(\omega)
	& = 2 \sum_{i}^\text{occ} \sum_{x} \frac{[pi|x]^2}{\omega - \e{i} + \Om{x} - i \eta},
	\\
\label{eq:SigCp}
	\SigCp{p}(\omega)
	& = 2 \sum_{a}^\text{virt} \sum_{x} \frac{[pa|x]^2}{\omega - \e{a} - \Om{x} + i \eta},
\end{align}
\end{subequations}
where $\eta$ is a positive infinitesimal.
The screened two-electron integrals
\begin{equation}
	[pq|x] = \sum_{ia} (pq|ia) (\bX+\bY)_{ia}^{x}
\end{equation}
are obtained via the contraction of the bare two-electron integrals \cite{Gill_1994} $(pq|rs)$ and the transition densities $(\bX+\bY)_{ia}^{x}$ originating from a random phase approximation (RPA) calculation \cite{Casida_1995, Dreuw_2005}
\begin{equation}
\label{eq:LR}
	\begin{pmatrix}
		\bA	&	\bB	\\
		\bB	&	\bA	\\
	\end{pmatrix}
	\begin{pmatrix}
		\bX	\\
		\bY	\\
	\end{pmatrix}
	=
	\bOm
	\begin{pmatrix}
		\boldsymbol{1}	&	\boldsymbol{0}	\\
		\boldsymbol{0}	&	\boldsymbol{-1}	\\
	\end{pmatrix}
	\begin{pmatrix}
		\bX	\\
		\bY	\\
	\end{pmatrix},
\end{equation}
with
\begin{align}
\label{eq:RPA}
	A_{ia,jb} & = \delta_{ij} \delta_{ab} (\epsilon_a - \epsilon_i) + 2 (ia|jb),
	&
	B_{ia,jb} & = 2 (ia|bj),
\end{align}
and $\delta_{pq}$ is the Kronecker delta. \cite{NISTbook}
\titou{The one-electron energies $\epsilon_p$ in \eqref{eq:SigCh}, \eqref{eq:SigCp} and \eqref{eq:RPA} are either the HF or the GW quasiparticle energies.}
Equation \eqref{eq:LR} also provides the neutral excitation energies $\Om{x}$.

In practice, there exist two ways of determining the {\GOWO} QP energies. \cite{Hybertsen_1985a, vanSetten_2013}
In its ``graphical'' version, they are provided by one of the many solutions of the (non-linear) QP equation
\begin{equation}
\label{eq:QP-G0W0}
	\omega =  \eHF{p} + \Re[\SigC{p}(\omega)].
\end{equation}
In this case, special care has to be taken in order to select the ``right'' solution, known as the QP solution.
In particular, it is usually worth calculating its renormalization weight (or factor), $\Z{p}(\eHF{p})$, where
\begin{equation}
\label{eq:Z}
	\Z{p}(\omega) = \qty[ 1 - \pdv{\Re[\SigC{p}(\omega)]}{\omega} ]^{-1}.
\end{equation}
Because of sum rules, \cite{Martin_1959, Baym_1961, Baym_1962, vonBarth_1996} the other solutions, known as satellites, share the remaining weight.
In a well-behaved case (belonging to the weakly correlated regime), the QP weight is much larger than the sum of the satellite weights, and of the order of $0.7$-$0.9$.

Within the linearized version of {\GOWO}, one assumes that
\begin{equation}
	\label{eq:SigC-lin}
	\SigC{p}(\omega) \approx \SigC{p}(\eHF{p}) + (\omega - \eHF{p}) \left. \pdv{\SigC{p}(\omega)}{\omega} \right|_{\omega = \eHF{p}},
\end{equation}
that is, the self-energy behaves linearly in the vicinity of $\omega = \eHF{p}$.
Substituting \eqref{eq:SigC-lin} into \eqref{eq:QP-G0W0} yields
\begin{equation}
\label{eq:QP-G0W0-lin}
	\eGOWO{p} = \eHF{p} + \Z{p}(\eHF{p}) \Re[\SigC{p}(\eHF{p})].
\end{equation}
Unless otherwise stated, in the remaining of this paper, the {\GOWO} QP energies are determined via the linearized method.

In the case of {\evGW}, the QP energy, $\eGW{p}$, are obtained via Eq.~\eqref{eq:QP-G0W0}, which has to be solved self-consistently due to the QP energy dependence of the self-energy [see Eq.~\eqref{eq:SigC}]. \cite{Hybertsen_1986, Shishkin_2007, Blase_2011, Faber_2011} 
At least in the weakly correlated regime where a clear QP solution exists, we believe that, within {\evGW}, the self-consistent algorithm should select the solution of the QP equation \eqref{eq:QP-G0W0} with the largest renormalization weight $\Z{p}(\eGW{p})$.
In order to avoid convergence issues, we have used the DIIS convergence accelerator technique proposed by Pulay. \cite{Pulay_1980, Pulay_1982}
\titou{Details about our implementation of DIIS for evGW can be found in the Appendix.}
Moreover, throughout this paper, we have set $\eta = 0$.

\begin{figure}
	\includegraphics[width=0.6\linewidth]{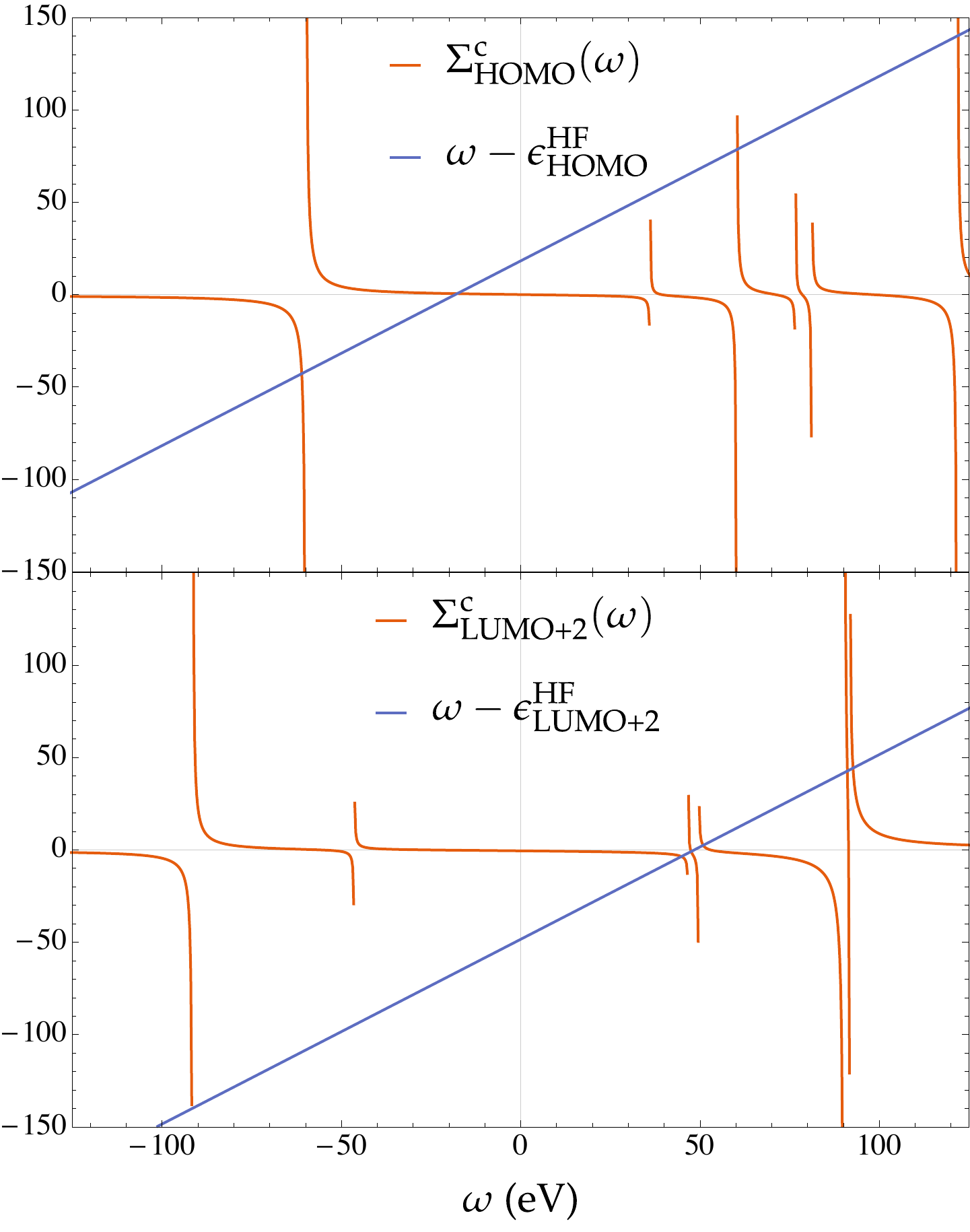}
	\caption{
	\label{fig:H2-QPvsOm}
	$\SigC{\HOMO}(\omega)$ and $\SigC{\LUMO+2}(\omega)$ (in eV) as functions of the frequency $\omega$ obtained at the {\evGW}@HF/6-31G level for \ce{H2} at $\RH = 1.0$ bohr.
	The solutions of the QP equation are given by the intersection of the \titou{orange} and blue curves.}
\end{figure}

\section{Results}
\subsection{Virtual orbitals}
As a first example, we consider the hydrogen molecule \ce{H2} in a relatively small gaussian basis set (6-31G) in order to be able 
to study easily the entire orbital energy spectrum.
Although the number of irregularities/discontinuities as well as their locations may vary with the basis set, the conclusions we are going to draw here are general.

Figure \ref{fig:H2} reports three key quantities as functions of the internuclear distance $\RH$ for various orbitals at the {\GOWO} and the self-consistent {\evGW} levels: 
i) the QP energies [$\eGOWO{p}$ or $\eGW{p}$],
ii) the correlation part of the self-energy [$\SigC{p}(\eHF{p})$ or $\SigC{p}(\eGW{p})$], and 
iii) the renormalization factor/weight [$\Z{p}(\eHF{p})$ or $\Z{p}(\eGW{p})$].

\subsubsection{{\GOWO}}
Let us first consider the results of the {\GOWO} calculations reported in the top row of Fig.~\ref{fig:H2}.
Looking at the curves of $\eGOWO{p}$ as a function of $\RH$ (top left graph of Fig.~\ref{fig:H2}), one notices obvious irregularities 
in the LUMO+2 around $\RH = 1.0$ bohr and in the LUMO+1 around $\RH = 2.1$ bohr.
For information, the experimental equilibrium geometry of \ce{H2} is around $\RH = 1.4$ bohr. \cite{HerzbergBook}
These irregularities are unphysical, and occur in correspondence with a series of poles in $\SigC{\LUMO+1}$ and $\SigC{\LUMO+2}$ (see top center graph of Fig.~\ref{fig:H2}).
For example, one can notice two poles in $\SigC{\LUMO+2}$ just before and after $\RH = 1.0$ bohr, giving birth to three branches.
The origin of the irregularities in $\e{\LUMO+1}$ and $\e{\LUMO+2}$ can, therefore, be traced back to the wrong assumption that $\SigC{\LUMO+1}(\omega)$ and $\SigC{\LUMO+2}(\omega)$ are linear functions of $\omega$ in the vicinity of, respectively, $\omega = \eHF{\LUMO+1}$ and $\omega = \eHF{\LUMO+2}$ [see Eq.~\eqref{eq:SigC-lin}].

However, despite the divergencies in the self-energy, the QP energies $\eGOWO{\LUMO+1}$ and $\eGOWO{\LUMO+2}$ remain finite thanks to a rapid decrease 
of the renormalization factor at the $\RH$ values for which the self-energy diverges [see Eq.~\eqref{eq:QP-G0W0} and top right graph of Fig.~\ref{fig:H2}].
For example, note that $\Z{\LUMO+2}$ reaches exactly zero at the pole locations.
A very similar scenario unfolds for the LUMO+1, except that a single pole is present in $\SigC{\LUMO+1}$.

Let us analyze this point further.
Since the self-energy behaves as $\SigC{p} \sim \delta^{-1}$ (with $\delta \to 0$) in the vicinity of a singularity, one can easily show that $\Z{p} \sim (1+\delta^{-2})^{-1} \sim \delta^{2}$, which yields $\eGOWO{p} \sim \eHF{p} +\delta$.
In plain words, $\eGOWO{p}$ remains finite near the poles of the self-energy thanks to the linearization of the QP equation [see Eq.~\eqref{eq:QP-G0W0}].
It also evidences that, at the pole locations (i.e.~$\delta = 0$), we have $\eGOWO{p} = \eHF{p}$, i.e., by construction the QP energy is forced to remain equal to the zeroth-order energy.
This is nicely illustrated in Fig.~\ref{fig:H2-zoom}, where we have plotted the HF orbital energies (dotted lines) as well as the {\GOWO} QP energies (solid lines) around the two ``problematic'' internuclear distances.
The behavior of $\eGOWO{\LUMO+1}$ (solid orange line on the right panel of Fig.~\ref{fig:H2-zoom}) is particularly instructive and shows that the {\GOWO} QP energies can have an erratic behavior near the poles of the self-energy.

It is interesting to investigate further the origin of these poles.
As evidenced by Eq.~\eqref{eq:SigC}, for a calculation involving $2n$ electrons and $N$ basis functions, the self-energy has exactly $n N (N-n)$ poles originating from the combination of the $N$ poles of the Green function $G$ (at frequencies $\e{p}$) and the $n(N-n)$ poles of the screened Coulomb interaction $W$ (at the RPA singlet excitations $\Om{x}$).
For example, at $\RH = 2.11$ bohr, the combination of $\eHF{\LUMO} = 3.83$ eV and the HOMO-LUMO-dominated first neutral excitation energy $\Om{1} = 22.24$ eV are equal to the LUMO+1 energy $\eGOWO{\LUMO+1} = 26.07$ eV.
Around $\RH=1.0$ bohr, the two poles of $\SigC{\LUMO+1}$ are due to the following accidental equalities: $\eGOWO{\LUMO+1} = \eHF{\LUMO} + \Om{2}$, and $\eGOWO{\LUMO+1} = \eHF{\LUMO+1} + \Om{1}$.
Because the number of poles in $G$ and $W$ \titou{(at the non-interacting or HF level)} are both proportional to $N$, these spurious poles in the self-energy become more and more frequent for larger gaussian basis sets.
For virtual orbitals, the higher in energy the orbital is, the earlier the singularities seem to appear.

Finally, the irregularities in the {\GOWO} QP energies as a function of $\RH$ can also be understood as follows.
Since within {\GOWO} only one pole of $G$ is calculated, i.e., the QP energy, all the satellite poles are discarded.
Mixing between QP and satellites poles, which is important when they are close to each other, hence, is not considered.
This situation can be compared to the lack of mixing between single and double excitations in
adiabatic time-dependent density-functional theory and the
Bethe-Salpeter equation \cite{Maitra_2004, Cave_2004, Romaniello_2009_JCP, Sangalli_2011} (see also Refs.~\citenum{Li_2014, Ou_2015, Zhang_2015, Parker_2016}).

\subsubsection{{\evGW}}
 Within partially self-consistent schemes, the presence of poles in the self-energy at a frequency similar to a QP energy has more dramatic consequences.
The results for \ce{H2} at the {\evGW}@HF/6-31G level are reported in the bottom row of Fig.~\ref{fig:H2}.
Around $\RH = 1.0$ bohr, we observe that, for the LUMO+2, one can fall onto three distinct solutions depending on the algorithm one relies on to solve self-consistently the QP equation (see bottom left graph of Fig.~\ref{fig:H2}).
In order to obtain each of the three possible solutions in the vicinity of $\RH = 1.0$ bohr, we have run various sets of calculations using different starting values for the QP energies and sizes of the DIIS space.
In particular, we clearly see that each of these solutions yield a distinct energy separated by several electron volts (see zoom in Fig.~\ref{fig:H2-zoom}), and each of them is associated with a well-defined branch of the self-energy, as shown by the center graph in the bottom row of Fig.~\ref{fig:H2}.
For convenience, the intermediate (center) branch is presented in lighter green in Figs.~\ref{fig:H2} and \ref{fig:H2-zoom}, while the left and right branches are depicted in darker green.
Interestingly, the {\evGW} iterations are able to ``push'' the QP solution away from the poles of the self-energy, which explains why the renormalization factor is never exactly equal to zero (see bottom right graph of Fig.~\ref{fig:H2}).
However, one cannot go smoothly from one branch to another, and each switch between solutions implies a significant energetic discontinuity.
Moreover, we observe ``ripple'' effects in other virtual orbitals: a discontinuity in one of the QP energies induces (smaller) discontinuities in the others. 
This is a direct consequence of the global energy dependence of the self-energy [see Eq.~\eqref{eq:SigC}], and is evidenced on the left graph in the bottom row of Fig.~\ref{fig:H2} around $\RH = 2.1$ bohr.

The main observation of the present study is that each branch of the self-energy is associated with a distinct QP solution.
We clearly see that, when one goes from one branch to another, there is a transfer of weight between the QP and one of the satellites, 
which becomes the QP on the new branch. \cite{Loos_2018}
As opposed to the strongly correlated regime where the QP picture breaks down, i.e., there is no clear QP, 
here there is alway a clear QP except at the vicinity of the poles where the weight transfer occurs.
As for {\GOWO}, this sudden transfer is caused by the artificial removal of the satellite poles. However, in the {\evGW} results the problem is amplified by the self-consistency. 
We expect that keeping the full frequency dependence of the self-energy would solve this problem.

It is also important to mention that the self-consistent algorithm is fairly robust as it rarely selects a solution with a renormalization weight lower than $0.5$, as shown by the center graph in the bottom row of Fig.~\ref{fig:H2}.
In other words, when the renormalization factor of the QP solution becomes too small, the self-consistent algorithm switches naturally to a different solution.
From a technical point of view, around the poles of the self-energy, it is particularly challenging to converge self-consistent calculations, and we heavily relied on DIIS to avoid such difficulties.
We note that an alternative \textit{ad hoc} approach to stabilize such self-consistent calculations is to increase the value of the positive infinitesimal $\eta$.

Figure \ref{fig:H2-QPvsOm} shows the correlation part of the self-energy for the HOMO and LUMO+2 orbitals as a function of $\omega$ (\titou{orange} curves) obtained at the self-consistent {\evGW}@HF/6-31G level for \ce{H2} with $\RH = 1.0$ bohr.
The solutions of the QP equation \eqref{eq:QP-G0W0} are given by the intersections of the \titou{orange} and blue curves.
On the one hand, in the case of the HOMO, we have an unambiguous QP solution (at $\omega \approx -20$ eV) which is well separated from the other solutions.
In this case, one can anticipate a large value of the renormalization factor $\Z{\HOMO}$ as the self-energy is flat around the intersection of the two curves.
On the other hand, for the LUMO+2, we see three solutions of the QP equation very close in energy from each other around $\omega = 50$ eV.
In this particular case, there is no well-defined QP peak as each solution has a fairly small weight.
Therefore, one may anticipate multiple solution issues when a solution of the QP equation is close to a pole of the self-energy.

Finally, we note that the multiple solutions discussed here are those of the QP equation, i.e., \emph{multiple} QP poles associated to a \emph{single} Green function. 
This is different from the multiple solutions discussed in Refs.~\citenum{Kozik_2014, Stan_2015, Rossi_2015, Tarantino_2017, Schaefer_2013, Schaefer_2016, Gunnarsson_2017}, in which it is shown that, in general, the nonlinear Dyson equation admits \emph{multiple} Green functions, which can be physical but also unphysical.

\begin{figure*}
	\includegraphics[width=\linewidth]{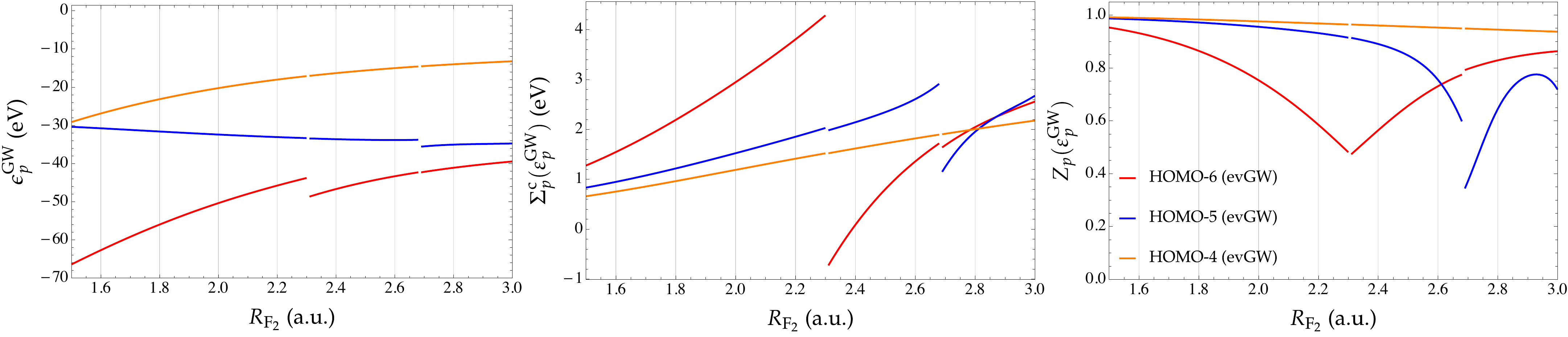}
	\caption{
	\label{fig:F2}
	QP energies (left), correlation part of the self-energy (center) and renormalization factor (right) as functions of the internuclear distance $\RF$ for various  occupied orbitals of \ce{F2} at the {\evGW}@HF/STO-3G level.}
\end{figure*}

\subsection{Occupied orbitals}
So far, we have seen that multiple solutions seem to only appear for virtual orbitals (LUMO excluded).
However, we will show here that it can also happen in occupied orbitals.
We take as an example the fluorine molecule (\ce{F2}) in a minimal basis set (STO-3G), and perform {\evGW}@HF calculations within the frozen-core approximation, that is, we do not update the orbital energies associated with the core orbitals.
Figure \ref{fig:F2} shows the behavior (as a function of the distance between the two fluorine atoms $\RF$) of the same quantities as in Fig.~\ref{fig:H2} but for some of the occupied orbitals of \ce{F2} (HOMO-6, HOMO-5 and HOMO-4).
Similarly to the case of \ce{H2} discussed in the previous section, we see discontinuities in the QP energies around $\RF = 2.3$ bohr (for the HOMO-6) and $\RF = 2.7$ bohr (for the HOMO-5). 
For information, the experimental equilibrium geometry of \ce{F2} is $\RF = 2.668$ bohr, which evidences that the second discontinuity is extremely close to the experimental geometry.
Let us mention here that we have not found any discontinuity in the HOMO orbital.
The case of the frontier orbitals will be discussed below.
For \ce{F2}, here again, we clearly observe ripple effects on other occupied orbitals. 
Similarly to virtual orbitals, we have found that the lower in energy the orbital is, the earlier the singularities seem to appear.

\begin{figure}
	\includegraphics[width=0.6\linewidth]{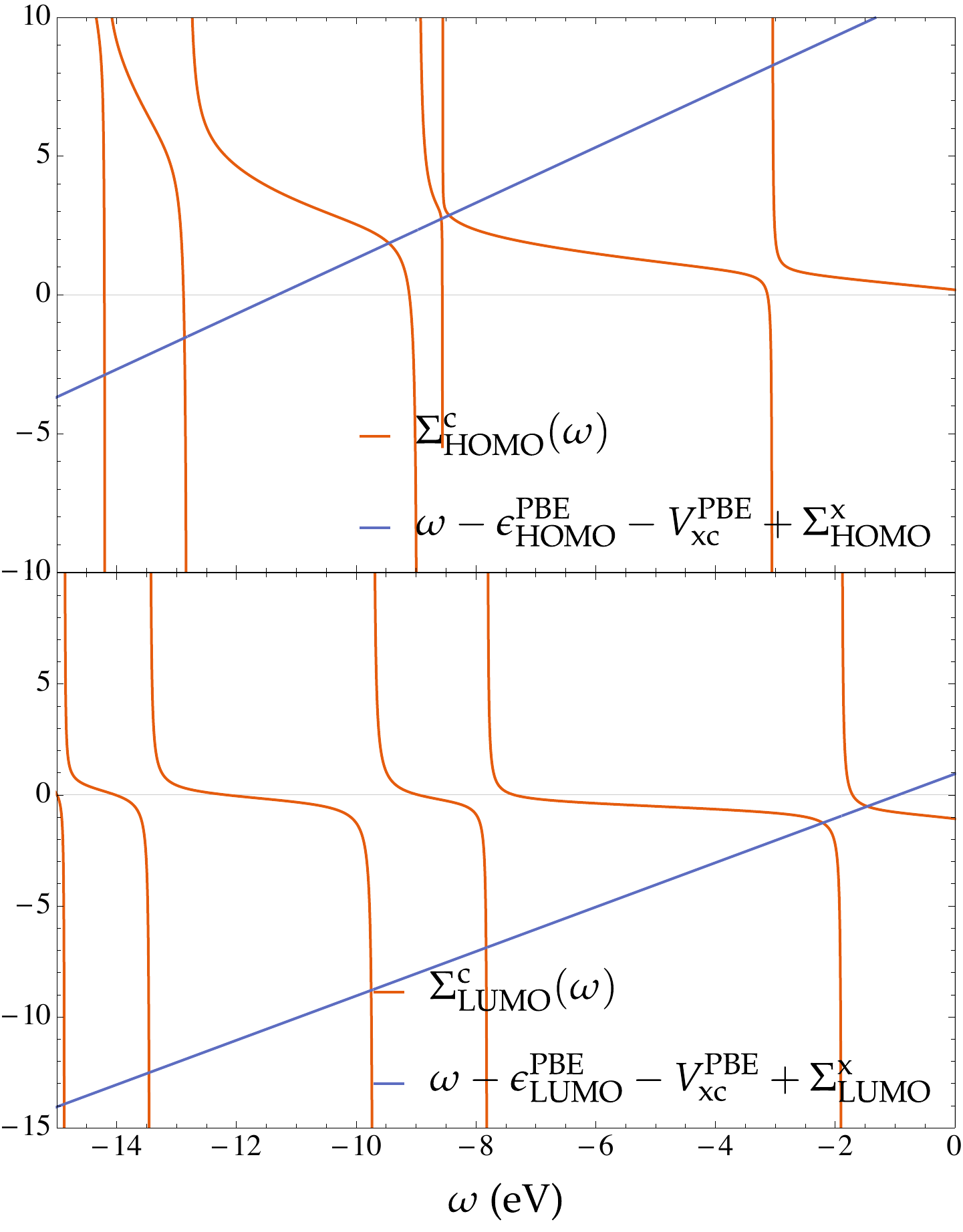}
	\caption{
	\label{fig:BeO}
	$\SigC{\HOMO}(\omega)$ and $\SigC{\LUMO}(\omega)$ (in eV) as functions of the frequency $\omega$ obtained at the {\GOWO}@PBE/cc-pVDZ level for \ce{BeO} at its experimental geometry. \cite{HerzbergBook}
	The solutions of the QP equations are given by the intersection of the \titou{orange} and blue curves.}
\end{figure}

\subsection{Frontier orbitals}
Before concluding, we would like to know, whether or not, this multisolution behavior can potentially appear in frontier orbitals.
This is an important point to discuss as these orbitals are directly related to the ionization potential and the electron affinity, hence to the gap.

Let us take the HOMO orbital as an example.
A similar rationale holds for the LUMO orbital.
According to the expression of the hole and particle parts of the self-energy given in Eqs.~\eqref{eq:SigCh} and \eqref{eq:SigCp} respectively, $\SigC{\HOMO}(\omega)$ has poles at $\omega = \e{i} - \Om{x}$ and $\omega = \e{a} + \Om{x}$ with $\Om{x} > 0$.
Evaluating the self-energy at $\omega = \e{\HOMO}$ would yield $\e{\HOMO} - \e{i} = - \Om{x}$ and $\e{\HOMO} - \e{a} = + \Om{x}$, which is in clear contradiction with the assumption that $\Om{x} > 0$.
Therefore, the self-energy is never singular at $\omega=\eHOMO$ and $\omega=\eLUMO$ and the linearized {\GOWO} equations can be solved without any problem for the frontier orbitals.
This is true for any $G_0$, that is, it does not depend on the starting point.
As can be seen from Eqs.~\eqref{eq:SigCh} and \eqref{eq:SigCp}, the two poles of the self-energy closest to the Fermi level are located at $\omega = \eHOMO - \Om{1}$ and $\omega = \eLUMO + \Om{1}$.
As a consequence, there is a region equal to $\eHOMO - \eLUMO + 2\Om{1}$ around the Fermi level in which the self-energy does not have poles.
Because $\Om{1} \approx \eHOMO - \eLUMO = \Egap$, this region is approximately equal to $3\Egap$.

For ``graphical'' {\GOWO}, the solution might lie outside this range, even for the frontier orbitals.
This can happen when $\Egap$ is much smaller than the true GW gap.
In particular, this could occur for graphical {\GOWO} on top of a Kohn-Sham starting point, which is known to yield gaps that are (much) smaller than GW gaps.
Within graphical {\GOWO}, multiple solution issues for the HOMO have been reported by van Setten and coworkers \cite{vanSetten_2015, Maggio_2017} in several systems (\ce{LiH}, \ce{BN}, \ce{BeO} and \ce{O3}).
In their calculations, they employed PBE orbital energies \cite{Perdew_1996} as starting point, and this type of functionals is well known to drastically underestimate $\Egap$. \cite{ParrBook}

As an example, we have computed, within the frozen-core approximation, $\SigC{\HOMO}(\omega)$ and $\SigC{\LUMO}(\omega)$ as functions of $\omega$ at the {\GOWO}@PBE/cc-pVDZ level for beryllium monoxide (\ce{BeO}) at its experimental geometry (i.e.~$\RBeO = 2.515$ bohr). \cite{HerzbergBook}
These calculations have been performed with MOLGW. \cite{Bruneval_2016}
The results are gathered in Fig.~\ref{fig:BeO}, where one clearly sees that multiple solutions appear for both the HOMO and LUMO orbitals.
Note that performing the same set of calculations with a HF starting point yields a perfectly unambiguous single QP solution.
For this system, PBE is a particularly bad starting point for a GW calculation with a HOMO-LUMO gap equal to $1.35$ eV.
Using the same basis set, HF yields a gap of $8.96$ eV, while {\GOWO}@HF and {\GOWO}@PBE yields $7.54$ and $5.60$ eV.
The same observations can be made for the other systems reported as problematic by van Setten and coworkers. \cite{vanSetten_2015, Maggio_2017}
As a general rule, it is known that HF is usually a better starting point for GW in small molecular systems. \cite{Blase_2011, Bruneval_2013, Loos_2018, Langre_2018}
For larger systems, hybrid functionals \cite{Becke_1993} might be the ideal compromise, thanks to the increase of the HOMO-LUMO gap via the addition of (exact) HF exchange. \cite{Bruneval_2013, Boulanger_2014, Bruneval_2015, Jacquemin_2015, Gui_2018}

\section{Concluding remarks}
The GW approximation of many-body perturbation theory has been highly successful at predicting the electronic properties of solids and molecules. \cite{Onida_2002, Aryasetiawan_1998, Reining_2017}
However, it is also known to be inadequate to model strongly correlated systems. \cite{Romaniello_2009, Romaniello_2012, DiSabatino_2015, DiSabatino_2016, Tarantino_2017}
Here, we have found severe shortcomings of two widely-used variants of GW in the weakly correlated regime.
We have evidenced that one can hit multiple solution issues within {\GOWO} and {\evGW} due to the location of the QP solution near poles of the self-energy.
Within linearized {\GOWO}, this implies irregularities in key experimentally-measurable quantities of simple diatomics, while, at the partially self-consistent {\evGW} level, discontinues arise. 
Because the RPA correlation energy \cite{Casida_1995, Dahlen_2006, Furche_2008, Bruneval_2016} and the Bethe-Salpeter excitation energies \cite{Strinati_1988, Leng_2016, Blase_2018} directly dependent on the QP energies, these types of discontinuities are also present in these quantities, hence in the energy surfaces of ground and excited states.
Illustrative examples can be found in our previous study. \cite{Loos_2018}
\titou{We believe that such discontinuities would not exist within a fully self-consistent scheme where one does not iterate the QP energies but the one-body Green's function and therefore takes into account each QP peak as well as its satellites at every iteration.}
Obviously, this latter point deserves further investigations.
However, if confirmed, this would be a strong argument in favor of fully self-consistent schemes.
Also, for extended systems, these issues might be mitigated by the plasmon modes that dominate the high-energy spectrum of the screened Coulomb interaction.
\titou{The results of this work will be useful for self-consistent GW calculations of dynamical phenomena, i.e., with nuclear motion.}

\titou{We are currently exploring different routes in order to remove these unphysical features.
Pad\'e resummation technique could be of great interest \cite{Pavlyukh_2017} for such purpose.
However, other techniques might be successful at alleviating this issue. 
For example, one could i) impose a larger offset from the real axis (i.e.~increasing the value of $\eta$), ii) favor, in the case of small systems, a HF starting point in order to avoid small HOMO-LUMO gaps, or iii) rely, for larger systems, on hybrid functionals including a significant fraction of HF exchange. 
Also, regularization techniques, such as the one developed for orbital-optimized second-order M{\o}ller-Plesset perturbation theory, could be pragmatic and efficient way of removing such discontinuities. \cite{Lee_2018}}
 
\begin{acknowledgement}
PFL would like to thank Xavier Blase and Fabien Bruneval for valuable discussions.
The authors would like to thank Valerio Olevano for stimulating discussions during his sabbatical stay at IRSAMC.
MV thanks \textit{Universit\'e Paul Sabatier} (Toulouse, France) for a PhD scholarship.
\end{acknowledgement}

\section*{Appendix: DIIS implementation for GW}
\titou{DIIS (standing for ``direct inversion of the iterative subspace'') is an extrapolation technique introduced by Pulay in 1980 \cite{Pulay_1980, Pulay_1982} in order to speed up the convergence of self-consistent HF calculations.
The DIIS implementation for the evGW method is rather straightforward and reminiscent of the coupled cluster (CC) implementation. \cite{Scuseria_1986}
Within evGW, at iteration $n$, DIIS provides a set of normalized weight $\boldsymbol{w}$ in order to extrapolate the current values of the QP energies $\beGnWn{n}$ based on the $\nDIIS = \min (n-1,\maxDIIS)$ previous values, i.e.,
\begin{equation}
	\beGnWn{n} = \sum_{m=1}^{\nDIIS} w_m \beGnWn{n-m},
\end{equation}
where $\maxDIIS$ is a user-defined parameter setting the maximum size of the DIIS space.
This procedure only requires to store the QP energies $\beGnWn{n}$ at each iteration.
The DIIS extrapolation technique relies on the fact that, at convergence, $\bdeGnWn{n-1} = \beGnWn{n-1} - \beGnWn{n-2} = \bm{0}$.
Consequently, the weights are obtained by solving the linear system $\boldsymbol{A} \boldsymbol{w} = \boldsymbol{b}$, where
\begin{align}
	\boldsymbol{A} = & 
	\begin{pmatrix}
		(\bdeGnWn{n-1})^\dag \, \bdeGnWn{n-1}	&	\cdots	&	(\bdeGnWn{n-\nDIIS})^\dag \, \bdeGnWn{n-1}	&	-1			\\
		\vdots	&	\ddots	& \vdots	& \vdots	\\
		(\bdeGnWn{n-1})^\dag \, \bdeGnWn{n-\nDIIS}  	&	\cdots	&	(\bdeGnWn{n-\nDIIS})^\dag \, \bdeGnWn{n-\nDIIS}	&	-1			\\
		-1		&	\cdots	&	-1	&	0	\\
	\end{pmatrix},
	\\
	\boldsymbol{b} = & 
	\begin{pmatrix}
		0		\\
		\vdots	\\
		0		\\
		-1		\\
	\end{pmatrix}.
\end{align}
When the linear system becomes ill-conditioned, we reset $\nDIIS = 0$ and restart the DIIS extrapolation procedure.
For $\maxDIIS = 2$, the present algorithm can be seen as an optimal linear mixing strategy, as usually implemented in other softwares. \cite{Kaplan_2016, Caruso_2013a}
For qsGW, we have found that extrapolating the self-energy similarly to what is done for the Fock matrix in HF or KS methods is particularly efficient. \cite{Pulay_1980, Pulay_1982}}

\providecommand{\latin}[1]{#1}
\makeatletter
\providecommand{\doi}
  {\begingroup\let\do\@makeother\dospecials
  \catcode`\{=1 \catcode`\}=2 \doi@aux}
\providecommand{\doi@aux}[1]{\endgroup\texttt{#1}}
\makeatother
\providecommand*\mcitethebibliography{\thebibliography}
\csname @ifundefined\endcsname{endmcitethebibliography}
  {\let\endmcitethebibliography\endthebibliography}{}

\end{document}